\documentclass[a4paper]{jpconf}

\usepackage{graphicx,color}
\usepackage{amssymb}
\usepackage{amsmath}

\begin{document}
\title{Metal-insulator transition 
in the Edwards model}
\author{H.\ Fehske$^1$, S.\ Ejima$^{1}$, G. Wellein$^2$ and A. R. Bishop$^{3}$}
\address{$^{1}$Institut f{\"ur} Physik,
          Ernst-Moritz-Arndt-Universit{\"a}t Greifswald,
          17489 Greifswald,
          Germany\\$^{2}$RRZE, Friedrich-Alexander-Universit\"at Erlangen-N\"urnberg, 91058 Erlangen, Germany\\ 
          $^{3}$Los Alamos National Laboratory, Los Alamos, New Mexico 87545, U.S.
          }
\ead{<fehske,ejima>@physik.uni-greifswald.de}

\date{\today}

\begin{abstract}
To understand how charge transport is affected by a background 
medium and vice versa we study a two-channel transport model which 
captures this interplay via a novel, effective fermion-boson 
coupling. By means of (dynamical) DMRG we prove 
that this model exhibits a metal-insulator transition at half-filling, 
where the metal typifies a repulsive Luttinger liquid
and the insulator constitutes a charge density wave.
The quantum phase transition point is determined consistently from the 
calculated photoemission spectra, the scaling of the Luttinger liquid exponent,
the charge excitation gap, and the entanglement entropy. 
\end{abstract}
The way a system evolves from a metallic to an insulating state
is one of the most fundamental problems in solid state theory.
Electron-electron and electron-phonon interactions are the
driving forces behind metal-insulator transitions (MITs) in the
majority of cases. For example, the Mott-Hubbard MIT~\cite{Mot90} 
is caused by strong Coulomb correlations, whereas the Peierls
 MIT~\cite{Pe55} is triggered by the coupling to vibrational excitations of 
the crystal. Theoretically the MIT problem can be addressed by the 
investigation of generic Hamiltonians for interacting electrons and 
phonons such as Hubbard or Holstein models~\cite{Hu63}.
In one dimension (1D), these models exhibit 
a MIT at half-filling, where on the insulating side of 
the MIT a spin-density-wave (SDW) or a charge-density-wave     
(CDW) broken-symmetry ground state appears, respectively.
On the metallic side, near the MIT, charge transport then takes place 
within a strongly correlated ``background'' that anticipates 
the developing SDW, respectively CDW, order. Since the particles 
responsible for charge transport and the background order 
phenomena are the same, the problem is very complex. 

\begin{figure}[t]
   \begin{center}
    \includegraphics[width=.7\linewidth]{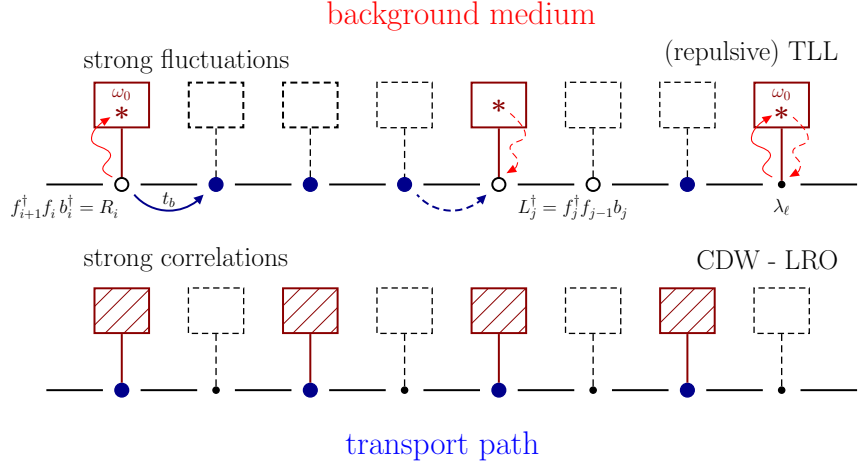}
   \end{center}
     \caption{The Edwards model~(\ref{model}) describes a very 
general situation: As a charge carrier ($\bullet$) moves 
along a 1D transport path it creates an excitation with energy $\omega_0$ 
($\ast$) in the background at the site it leaves or annihilates an 
existing excitation at the site it enters. The background 
medium may represent, e.g., a magnetically, orbitally or charge 
ordered lattice. One assumes that the (de)excitation 
of the background can be parameterized as a bosonic degree of 
freedom.  Any distortion of the background can heal 
by quantum fluctuations. Accordingly the $\lambda$-term 
allows for spontaneous boson creation and 
annihilation processes. }  
   \label{fig:model}
\end{figure}

A path forward might be the construction of simplified
transport models, which capture the basic mechanisms of 
quantum transport in a background medium in a rather effective way. 
Along this line a novel quantum transport model has been 
proposed recently~\cite{Ed06}, 
\begin{equation}
 {\cal H}= -t_b\sum_{\langle i, j \rangle} f_j^{\dagger}f_{i}^{}
  (b_i^{\dagger}+b_j^{})- \lambda\sum_i(b_i^{\dagger}+b_i^{})
              + \omega_0\sum_i b_i^{\dagger}b_i^{}\,.
\label{model}
\end{equation} 
This so-called Edwards model mimics the correlations 
inherent to a spinfull fermionic many-particle 
systems by a boson affected hopping of spinless particles 
(see Fig.~\ref{fig:model}). For the half-filled band case, 
the model describes a repulsive Tomonaga-Luttinger liquid (TLL), 
provided the excitations of the background are energetically 
inexpensive ($\omega_0 < \omega_{0,c}$) 
or will readily relax ($\lambda > \lambda_c(\omega_0)$).
This defines the fluctuation dominated regime.  
By contrast, strong background correlations, which will develop for 
large $\omega_0$ and small $\lambda\ll t_b$ tend to 
immobilize the charge carriers and may even drive 
a MIT by establishing CDW long-range order~\cite{WFAE08}.

In the present work, we employ density-matrix renormalization group (DMRG)
and dynamical DMRG methods~\cite{Wh92} to analyse the ground-state 
properties of the Edwards model and the charge carrier
dynamics for the limiting case of high-energy background fluctuations.   

Let us start with the discussion of the photoemission (PE) spectra. 
The single-particle spectral function probed by angle-resolved 
[inverse] PE reads
\begin{equation}
A(k,\omega)=A^-(k,\omega)+A^+(k,\omega)\,,\quad\mbox{with}\quad A^{\pm}(k,\omega)
=\sum_n|\langle\psi_n^{\pm}|f^{\pm}_k|\psi_0\rangle|^2\,
\delta[\omega\mp\omega^{\pm}]\,.
\label{spsfpm}
\end{equation}
Here $A^-(k,\omega)$ [$A^+(k,\omega)$] is associated with the emission
[injection] of an electron with wave vector $k$, 
i.e. $f^-_k=f^{}_k$ and $f^+_k=f^\dagger_k$.  $|\psi_0\rangle$ 
is the ground state of a $N$--site system in the $N_f$--particle sector,  
while $| \psi_n^{\pm}\rangle$ denote the $n$-th 
excited states in the $N_f\pm 1$-particle sectors 
with excitation energies $\omega_n^\pm=E_n^\pm-E_0$.
For the half-filled Edwards model we have $N_f=N/2$.

\begin{figure}[t]
\begin{center}
\includegraphics[width=0.95\linewidth]{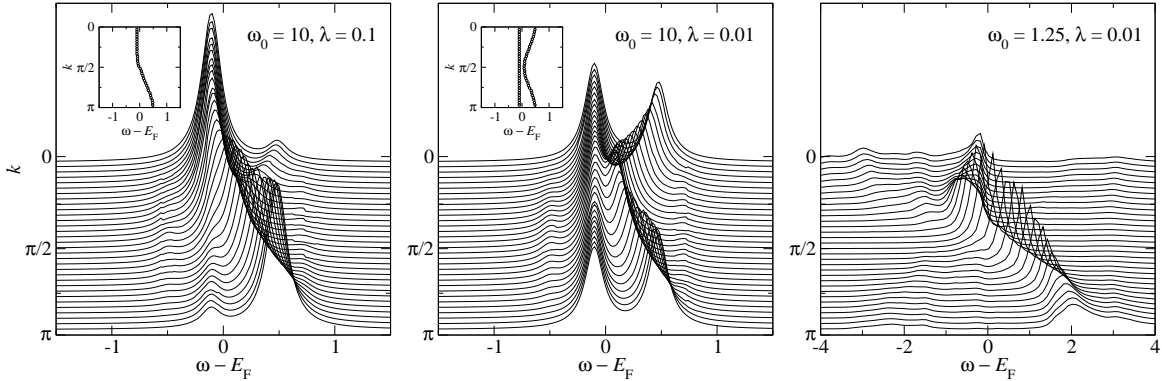}
\caption{Line-shape of the single-particle spectral $A(k,\omega)$
in the half-filled band sector of the 1D Edwards model. The insets shows the 
dispersion of the absorption/emission maximum. 
For the numerics,  we consider an $N=32$--site chain with open boundary conditions (BC) and
map a boson site, containing $2^{n_b}$ 
states, to $n_b$ pseudosites. We take up to 4 pseudosites,
keep $m=500$ density-matrix eigenstates, and use a 
broadening $\eta=0.1$. All energies are given in units of $t_b$.}  
   \label{fig:spsf}
  \end{center}
\end{figure}

Figure~\ref{fig:spsf} shows $A(k,\omega)$ for a stiff background, i.e.
 the distortions induced by particle hopping are energetically costly. 
In this regime the bosons will strongly affect particle transport: The 
quasiparticle mass is sizeably enhanced and a renormalized band structure 
appears but---if $\lambda$ is large enough---the system remains metallic,
as can be seen from the finite spectral weight at the Fermi
energy $E_F$ (left panel). As the system's ability for relaxation decreases,
i.e., at fixed $\omega_0$, $\lambda$ falls below a certain critical value,
a gap opens in the single-particle spectrum at $k_F=\pi/2$ (middle panel).
Evidently the system has become an insulator. We note  the internal feedback 
mechanism: The collective boson excitations originate from the motion of the 
charge carriers and have to persist long enough to finally inhibit particle transport,
thereby completely changing the nature of the many-particle ground state.
The collective boson-particle dynamics leads to an asymmetric band structure
for $k\leq k_F$ and $k\geq k_F$ (see inset). While the induced hole probed by PE can only move coherently 
by a six-step process with three bosons first excited and afterwards
consumed, an additional electron can easily move by a two-step process
even if strong CDW correlations exist in the background~\cite{WFAE08}.
We note that the (I)PE spectra exhibit weak signals around the bare boson energies
$\pm \omega_0$ (not shown). The interrelation of charge dynamics and background fluctuation becomes 
apparent again, if we decrease $\omega_0$ keeping $\lambda$ fixed (right panel).
Now the fluctuations overcome the correlations and the system returns
to a metallic state which is different in nature, however, from 
the state we started with: $A(k,\omega)$ shows sharp absorption features
near $k_F$ only and is ``overdamped'' at the Brillouin zone boundaries, where 
the spectrum is dominated by bosonic excitations.   
 
In order to determine more precisely the phase boundary between the metallic and insulating ground states, 
typifying a Tomonaga-Luttinger liquid (TLL) and a CDW, respectively, we analyse the limiting ($N\to\infty$) behaviour 
of the TLL charge exponent 
\begin{equation}
K_\rho=\pi\lim_{q\to 0}\frac{S_c(q)}{q}\,,\quad\mbox{with}\quad
S_c(q)=\tfrac{1}{N}\sum_{i,j}{\rm e}^{{\rm i}q (j-k)} \langle (n_j-\tfrac{1}{2})(n_k-\tfrac{1}{2})\rangle\,,\;\; q=\frac{2\pi}{N},
\label{krho}
\end{equation}
as well as those of the single-particle (charge) gap, 
$\Delta_c(N)=E_0^++E_0^--2E_0$,
and monitor the finite-size scaling of the entanglement entropy difference~\cite{Ni11}
\begin{equation}
\Delta S_N=S_N(N/2)-S_{N}(N/2-1)=-\frac{c^\ast}{3}\ln \cos\left[\frac{\pi}{N}\right]\,,
\label{delta_e_entropy}
\end{equation}
where $S_N(l)=-{\rm Tr} [\rho_l\ln \rho_l]= 
\tfrac{c^\ast}{3}\ln\left[ \tfrac{N}{\pi}\sin\Big(\frac{\pi l}{N}\Big)\right]+s_1$. 
We expect that the TLL charge exponent decreases from $K_{\rho}=1$, as $\lambda$ is lowered,
and finally reaches 1/2 at the MIT point, if the transition is of Kosterlitz-Thouless type~\cite{KT73,BMM98}.
The central charge $c^*$ should scale to unity  in the metallic TLL regime~\cite{CC04}.

\begin{figure}[t]
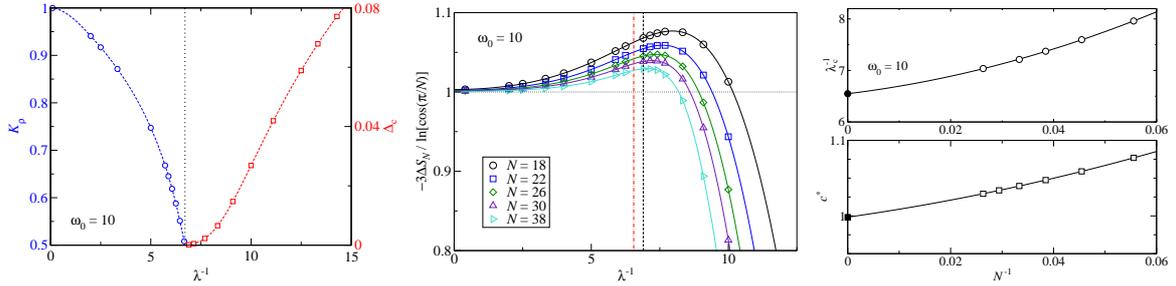

\begin{center}
\includegraphics[width=0.32\linewidth]{w10_only_lower_panel.eps}
\hspace*{0.1cm}\includegraphics[width=0.315\linewidth]{fig_ohne_inset.eps}
\hspace*{0.1cm}\includegraphics[width=0.29\linewidth]{fig_lambdac_maximum.eps}
\caption{Left panel: $N\to\infty$ extrapolated value of the TLL parameter
$K_{\rho}$, respectively of the charge gap $\Delta_c$, 
as a function of $\lambda^{-1}$ for $\omega_0=10$ 
(open BC). Middle panel: Entanglement entropy differences  
$\Delta S_N$ for different system sizes (periodic BC). The red 
dashed-dotted line gives the MIT transition point 
 in reasonable agreement with the value obtained from
$K_{\rho}$ (black dashed line). Right panels: Critical value
of $\lambda^{-1}_c$ (filled circle, top panel) 
and central charge $c^{\ast}\simeq 1$ (filled square, bottom panel),
both extrapolated from the maxima of $\Delta S_N$.
Here we use $m=2000$, $n_b=2$, and ensure a discarded weight less than $10^{-10}$.}  
   \label{fig:krhoetal}
  \end{center}

\end{figure} 

Figure~\ref{fig:krhoetal} demonstrates that the $N\to\infty$ extrapolated $K_{\rho}$ 
indeed becomes 1/2 at some critical
value, where $\lambda_c^{-1} (\omega_0=10)\simeq 5.89$, indicating the MIT. In the metallic phase
we find a repulsive particle interaction, $K_{\rho}\leq 1$. Our DMRG results point towards 
an  exponential opening of the charge gap entering the insulating state, which corroborates 
the Kosterlitz-Thouless transition scenario. Note that the CDW state of the Edwards model
is a few boson state, in contrast to the Peierls CDW phase of the Holstein model~\cite{WFAE08}.
That means the MIT in the Edwards model is driven by strong correlations, as for the Mott-Hubbard 
transition. To extract the central charge $c^*$ we use the entanglement entropy difference, 
Eq.~(\ref{delta_e_entropy}), rather than directly exploiting $S_N(l)$. For a model with spinless fermions
this is advantageous because we can work with  a fixed system size, thereby avoiding antiperiodic BC 
that give rise to complex phase factors~\cite{Ni11}.  As can be seen from the middle panel of Fig.~\ref{fig:krhoetal},
for $\lambda^{-1} <\lambda_c^{-1}$, the rescaled quantity $-3\Delta S_N/\ln [\cos (\pi/N)]$ extrapolates 
to unity as $N\to\infty$. This opens an alternative route to detect the MIT point.  We find that the $\lambda_c^{-1}(\omega_0)$ determined by extrapolating the maximum of $\Delta S_N$,  i.e. in a completely different manner, 
matches the critical value obtained from $K_{\rho}$ surprisingly well. Simultaneously, indeed $c^*\to 1$ (see right panel).

To summarise, we have studied the spectral and ground-state properties of the 1D Edwards fermion-boson transport
model by large-scale (dynamical) DMRG numerics. We showed that strong correlations within the background
medium will not only affect the charge-carrier's dynamics by enhancing the quasiparticle mass 
but may even trigger a metal-insulator quantum phase transition. The MIT transition point 
has been determined in good agreement both from the TLL charge exponent and the entanglement
entropy difference. We stress that to date only a very small number of microscopic model exists 
which have been rigorously shown to exhibit a MIT.

{\it Acknowledgements.} The authors would like to thank  A. Alvermann, 
D. M. Edwards, G. Hager and S. Nishimoto  for valuable discussions, and the RRZE
for providing computer resources.  This work is supported by DFG SFB 652 and, at Los Alamos, by CINT and the  USDoE.

\section*{References}
\bibliographystyle{iopart-num}
\providecommand{\newblock}{}

\end{document}